\documentclass[aps,prd,epsf,nopacs,amsmath,amssymb,graphics,floatfix,onecolumn,12pt]{revtex4-1}
\usepackage{graphicx}% Include figure files
\usepackage{dcolumn}% Align table columns on decimal point
\usepackage{bm}% bold math
\usepackage{color}
\usepackage{amsmath}
 \usepackage{relsize}
\usepackage{array}

\newcommand{\be}{\begin{equation}}
\newcommand{\ee}{\end{equation}}
\newcommand{\ba}{\begin{eqnarray}}
\newcommand{\ea}{\end{eqnarray}}
\newcommand{\ban}{\begin{eqnarray*}}
\newcommand{\ean}{\end{eqnarray*}}

\newcommand{\eq}[1]{(\ref{#1})}

\begin{document}
\title{Curious case of gravitational lensing by binary black holes: a tale of  
two photon spheres, new relativistic images and caustics.}

\author{$^1$Mandar Patil\footnote{Electronic address: mpatil@impan.pl},
$^2$Priti Mishra\footnote{Electronic address: 
pritimishra@hri.res.in},
$^3$D Narasimha\footnote{Electronic address: dna@tifr.res.in}
}

\affiliation{$^1$Institute of Mathematics of Polish Academy of Sciences, 
Sniadeckich 8, 00-956 Warsaw, Poland. \\
$^2$Harish-Chandra Research Institute, Chhatnag Road, Jhunsi, Allahabad-211019, 
India. \\
$^2$Homi Bhabha National Institute, Anushaktinagar, Mumbai 400094, India.
\\
$^3$Tata Institute of Fundamental Research, Homi Bhabha Road, Mumbai 400005, 
India. 
}

%\begin{abstract}

%\end{abstract}

%\pacs{04.20.Dw, 04.70.-s, 04.70.Bw}

\maketitle

\newpage

\section*{Abstract}
Binary black holes have been in limelight off late due to the detection of a
gravitational waves from coalescing compact binaries in the 
events GW150914 and GW151226. In this paper we study gravitational lensing by 
the binary black holes modelled as an equal mass Majumdar-Papapetrou di-hole 
metric and show that this system displays features that are quite unprecedented 
and absent in any other lensing configuration investigated so far in the 
literature. We restrict our attention to the light rays which move on the plane 
midway between the two identical black holes, which allows us to employ various 
techniques developed for the equatorial lensing in the spherically symmetric 
spacetimes. If distance between the two black holes is below a certain 
threshold value, then the system admits two photon spheres. As in the case of 
single black hole, infinitely many relativistic images are formed due to the 
light rays which turn back from the region outside the outer (unstable) photon 
sphere, all of which lie beyond a critical angular radius with respect to the 
lens. However in the presence of the inner (stable) photon sphere, the 
effective potential after admitting minimum turns upwards and blows up 
for the smaller values of radii and the light rays that enter the 
outer photon sphere can turn back, leading to the formation of a new set of 
infinitely many relativistic images, all of which lie below the critical 
radius from the lens mentioned above. As the distance between the two black 
hole is increased, two photon spheres approach one another, merge and 
eventually disappear. In the absence of the photon sphere, apart from the 
formation of a finite number of discrete relativistic images, the system 
remarkably admits a radial caustic, which has never been observed in the 
context of relativistic lensing before. Thus the system of binary black 
hole admits novel features both in the presence and absence of photon spheres. 
We discuss possible observational signatures and implications of the binary 
black hole lensing.

\newpage

\section{Introduction}
The first detection of gravitational waves was made recently by Advanced LIGO 
on September 14, 2015 validating the Einstein`s general theory of relativity 
\cite{det1}. It was soon followed by the second detection on December 26, 
2015. These events are referred to as GW150914 and GW151226 \cite{det2}.  In 
both the events the gravitational waves were generated by a pair of black holes 
which orbited one another with decaying orbits and eventually 
merged to form a single black hole. A gigantic amount of energy worth $3$ solar 
masses was emitted in the first event GW150914 within the timescale of few 
milli-seconds exceeding the integrated intensity of all stars 
in the observable universe by two orders of magnitude. These and many other 
observations related to binary black holes in electromagnetic band 
\cite{mac},\cite{oj287} suggest that the system consisting of 
two black holes in the close vicinity can host a wide variety 
of remarkable phenomenon that would be interesting both from 
the point of view of astrophysics and fundamental physics. In this paper we 
explore binary black holes from a different perspective. We study gravitational 
lensing by a pair of black holes and show that this system can exhibit novel 
features which are quite unprecedented and absent in any other gravitational 
lensing configuration studied so far. The analysis of the gravitational lensing
is also a timely from the point of view of Event Horizon Telescope and Gravity 
collaborations which would in near future take snapshots of supermassive 
black holes at the center of our and nearby 
galaxies \cite{Bambi},\cite{Johansen},\cite{Doeleman},\cite{Lu},\cite{Bartko},\cite{Cunha},\cite{Cunha1}.

The two body problem in general relativity is extremely difficult, in contrast 
with Newtonian gravity. This is essentially consequence of the fact that 
Einstein equations are complicated coupled non-liner partial differential 
equations that are difficult to deal with. There are no exact solutions to 
Einstein equations depicting binary black holes that could be applicable in the 
realistic astrophysical context. In this paper we work with the 
Majumdar-Papapetrou solution depicting multiple black holes in the equilibrium, 
which could be thought of as a simple toy model that might capture some 
of the features in a realistic scenario. Majumdar-Papapetrou metric is perhaps 
the simplest multi-black hole solution known so far. It was discovered 
independently by Majumdar and Papapetrou \cite{majumdar},\cite{papa} and later 
demonstrated to represent spacetime with multiple black holes 
\cite{hartle}. There were many interesting investigations that 
were carried out in the context of Majumdar-Papapetrou spacetime
\cite{Dettmann},\cite{Cornish},\cite{Dettmann1},\cite{shadow1},\cite{2phs},\cite{patil},\cite{shadow2}. 
We also note that the gravitational lensing in the context of the
full-fledged numerical relativity simulation of binary black hole merger was studied 
in \cite{Bohn}.

We focus on the case where Majumdar-Papapetrou metric represents two identical 
equal mass non-rotating black holes at rest with respect to one another at 
certain distance. The metric is static, has a rotational symmetry around 
the line joining two black holes and reflection symmetry about the plane that 
is midway between the two black holes, which intersects line joining two black 
holes orthogonally. We assume that the light source and observer are located on 
the symmetry plane and focus on light rays which are confined to move 
on this plane. This allows us to employ various techniques developed to study 
gravitational lensing on the equatorial plane of spherically symmetric 
spacetimes \cite{narasimha},\cite{virbhadra},\cite{bozza},\cite{bozza2},\cite{
sahu},\cite{ sahu2}. The central point of the symmetry plane which is also the 
point exactly midway between two black holes on the axis essentially acts as a 
gravitational lens. The photon sphere plays an important role in the 
relativistic gravitational lensing. Recently it was demonstrated \cite{2phs} 
that the number of photon spheres on the symmetry plane depends on the distance 
between the two black holes. If the distance between the two black holes is 
small, then the symmetry place exhibits two photon spheres, one of which is 
unstable and the other one is stable. As the distance is increased the two 
photon spheres approach each other, merge and disappear. So when the distance is 
larger than the certain critical threshold value, there are no photon 
spheres on the equatorial plane. The properties of relativistic images and caustics are radically 
different depending on whether two photon spheres are present and absent. 
The existence of the stable photon sphere in the context of the axially 
symmetric electro-vacuum spacetimes was studied in \cite{Dolan}. We would like 
to mention that what we refer to photon sphere is in fact a ring rather than being a sphere
since we restrict out attention on the plane midway between the black holes.

The structure of images formed due to the photons that turn back from the region
above the outer unstable photon sphere is same as in the case of 
Schwarzschild black hole \cite{virbhadra},\cite{bozza}. All the images lie 
beyond certain critical angular radius with respect to the lens. There are 
infinite images clubbed together close to this critical radius. 
In case of Schwarzschild black hole there is a dark region below the 
critical radius with no images, which is consequence of the fact that the light 
rays which enter the single unstable photon sphere never turn back and enter 
black hole. The situation is drastically different in case of the di-hole in 
the presence of second stable inner photon sphere. As we will show, the 
effective potential turns upward again and the photons that enter outer photon 
sphere can now turn back. This leads to the formation of new infinite set of 
relativistic images, all of which lie inside the critical radius mentioned 
above. The region which would have been dark in the case of a 
single black hole, is not dark in the presence of the two photon spheres. Again 
most of the images are crowded together close to the critical radius and few 
discrete images lie close to the lens. In the absence of the photon 
sphere a finite number of discrete relativistic images are formed. 
Interestingly the radial caustic is present in this case. This is the location 
where the map between image plane and source plane is degenerate. Relativistic 
radial caustic has never been reported before in any other investigation so 
far. Thus we demonstrate that new novel features arise both in the presence and 
absence of twin photon spheres. 

We note that the system similar to the one we investigated in this paper where 
there are two black holes and source which could be bright spot in the 
accretion disk or maybe in the gas cloud falling towards black holes may not be 
so hard to realize in the nature. Based on our analysis and expectation that we 
might capture some of the features in the realistic scenario, we anticipate 
that the gravitational lensing signature of such a system will be 
characteristic and peculiar. Thus gravitational lensing signal will allow 
us to locate a binary black hole system in the sky and thus reduce the number 
of parameters in the search for gravitational waves from binary black holes in 
the interferometric detectors such as LIGO, VIRGO and KAGRA using matched 
filtering. This suggests that our investigation could possibly have serious 
implications for gravitational wave data analysis.  

\section{Majumdar-Papapetrou di-hole metric and photon spheres}

In this section we describe the Majumdar-Papapetrou di-hole metric and null 
geodesics. In classical mechanics a system of particles at rest with charge 
equal to mass is in the equilibrium for arbitrary distance between the 
particles. Interestingly same is true in general relativity as well. A system 
of maximally charged non-rotating black holes are in equilibrium irrespective 
of the distance between them. The static electro-vacuum spacetime 
depicting this scenario is Majumdar-Papapetrou metric which is the simplest 
multi-black hole exact solution in general relativity and thus it has served as 
a toy-model for numerous analytical investigations as quoted in the introductory 
section.

\begin{figure}
\begin{center}
\includegraphics[width=0.8\textwidth]{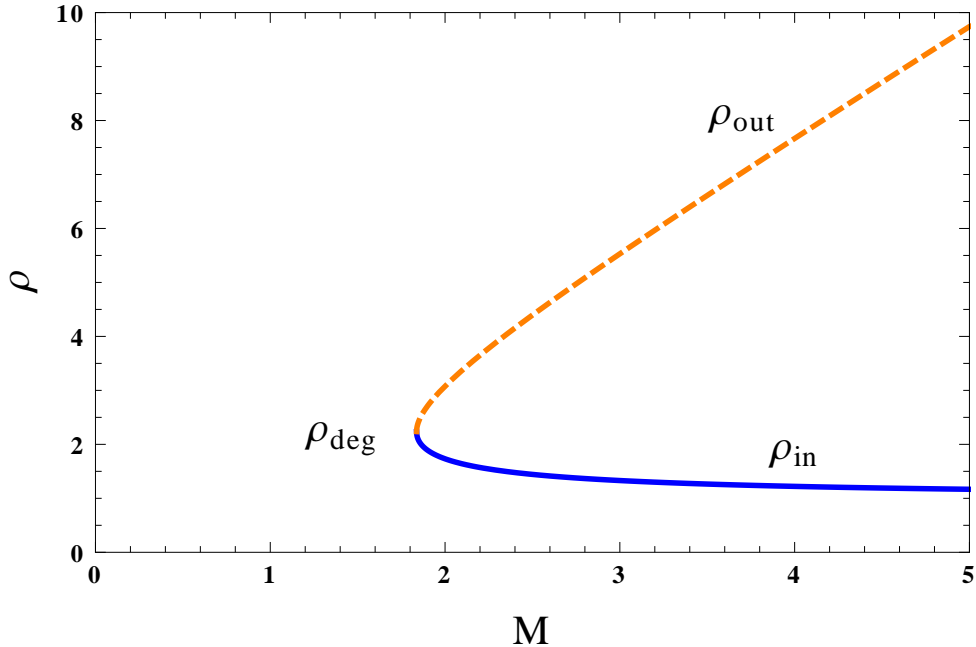}
\caption{\label{phs}
The radius of photon sphere $\rho$ is plotted against the mass to distance ratio
$M$. Below $M=M_{*}=\sqrt{\frac{27}{8}}$ there is no photon sphere. At 
$M=M_{*}$ we have a single degenerate photon sphere located at 
$\rho_{deg}=\sqrt{5}$. Above $M>M_{*}$ there are two photon spheres. The radius 
of outer photon sphere is denoted by $\rho_{out}$ and that of inner photon 
sphere is denoted by $\rho_{in}$.}
\end{center}
\end{figure}

In this paper we focus on the special case of the Majumdar-Papapetrou metric 
which represents two equal mass maximally charged non-rotating black holes in 
equilibrium at fixed distance between them. We call it a di-hole metric. The 
spacetime is static and asymptotically flat. It admits a rotational symmetry 
about the line joining two black holes and reflection symmetry about the plane 
that is midway between the black holes. In the cylindrical coordinate system 
$(t,\rho,\phi,z)$, which is well-adapted to the symmetries, the di-hole 
metric is given by
\begin{equation}
ds^2 = -\frac{dt^2}{U^2}+U^2(d\rho^2+\rho^2 d\phi^2+dz^2),
\end{equation}
where the metric function $U(\rho,z)$ is given by
\begin{equation}
   U(\rho,z) = 1 + \frac{M}{\sqrt{\rho^2+(z-a)^2}} + \frac{M}{\sqrt{\rho^2
   +(z+a)^2}}.
\end{equation}
Each of the black holes has mass $M$. The black holes are located at $z=+a$ and 
$z=-a$ on $z-$axis about which there is a rotational symmetry and the symmetry 
plane $z=0$ is located midway between the two black holes. In this coordinates 
black holes appear as points. We find it to convenient to work in the units 
where $a$ is set to unity and thus all dimensional quantities are appropriately 
expressed in units of $a$ which is half of the coordinate distance 
between the two black holes.  

We focus our attention on the null geodesics that are restricted 
to move on the symmetry plane $z=0$. Let $V$ be the four-velocity of the 
light ray. Using the standard techniques of dealing with null geodesics with the
help of conserved quantities associated with symmetries of metric and 
normalization of velocity \cite{sahu} we obtain the non-zero components of four-velocity as
\begin{eqnarray}
 &&V^{t}=\frac{U^2}{b} , \nonumber \\
 &&V^{\phi}=\pm \frac{1}{U^2\rho^2} ,\nonumber \\
 &&V^{\rho}=\pm \sqrt{\frac{1}{b^2}-\frac{1}{U^4\rho^2}} ,
 \label{vel}
\end{eqnarray}
where $b$ is the impact parameter, $\pm$ in the expression for 
$V^{\phi}$ stands for the photons moving clockwise and anti-clockwise 
respectively and $\pm$ in the expression for $V^{\rho}$ stands for the 
photons which move in radially outwards and inwards respectively. 
The function $U$ in the expression above is now 
\begin{equation}
   U(\rho) = 1 + \frac{2M}{\sqrt{\rho^2+1}} ,
\end{equation}
where $M$ is twice ratio of mass and distance between the black holes. The 
equation describing the radial motion of the light ray can be recast in the 
form 
\begin{equation}
 V^{\rho 2}+V_{eff}(\rho)=\frac{1}{b^2}~~;~~ V_{eff}(\rho)= \frac{1}{U^4\rho^2}.
\end{equation}
$V_{eff}$ is the effective potential for radial motion. It is quite useful to 
use the analogy of a particle moving in potential well in classical 
mechanics while dealing with the radial motion of photon. 

The location of the circular photon orbit i.e. photon sphere can be obtained by 
solving the equation 
$\frac{dV_{eff}}{d\rho}=0$ which is given by 
\begin{equation}
 \left(\rho^2+1\right)^{\frac{3}{2}}=2M \left(\rho^2-1\right).
\end{equation}
With the substitution $\eta^2=\rho^2+1$, the above equation can be cast into a 
cubic equation
\begin{equation}
 \eta^3-2M\eta^2+4M=0. 
\end{equation}
The existence or not of the photon spheres is determined by the 
discriminant of the cubic $\Delta=128M^2\left(M^2-27/8\right)$. The critical 
value of parameter $M$ for which discriminant is zero is given 
by 
\begin{equation}
 M_{*}=\sqrt{\frac{27}{8}}.
\end{equation}

When $M>M_{*}$, the discriminant of cubic is positive and two photon spheres are
present. We obtain location of the photon spheres by solving the cubic equation 
with Trigonometric method \cite{Dickson}. The location of the outer photon sphere 
denoted by $\rho_{out}$ is given by 
\begin{equation}
\rho_{out}= \sqrt{\frac{4M^2}{9}\left( 1+ 2\cos\left[\frac{1}{3} \cos^{-1}
\left(1-\frac{27}{4M^2}\right)\right] \right)^2-1  }.
\label{rout}
\end{equation}
The effective potential $V_{eff}$ admits a maximum at $\rho=\rho_{out}$ and 
hence the outer photon sphere is "unstable". The location of inner photon 
sphere denoted by $\rho_{in}$ is given by 
\begin{equation}
\rho_{in}= \sqrt{\frac{4M^2}{9}\left( 1- 2\sin\left[\frac{\pi}{6}-\frac{1}{3} 
\cos^{-1}\left(1-\frac{27}{4M^2}\right)\right] \right)^2-1  }.
\label{rin}
\end{equation}
Since the effective potential admits a minimum at $\rho=\rho_{in}$, the inner 
photon sphere is "stable". The location of photon spheres as a function of $M$ 
is depicted in Fig.\ref{phs}, which is same as Fig.4 in \cite{2phs}.

When $M=M_{*}$, two photon spheres coincide and we have a single degenerate 
photon sphere located at 
\begin{equation}
 \rho_{deg}=\sqrt{5}.
\end{equation}
Whereas in the case $M<M_{*}$, the discriminant of a cubic is negative and 
circular photon orbits are absent. 

As mentioned earlier $M$ is ratio of mass to the distance between the black 
holes. So the results obtained in this section imply that for a fixed mass, if 
the distance between the black holes is below certain critical value, two 
photon spheres are present on the symmetry plane $z=0$ midway between the black 
holes. As the black holes recede from another, two photon spheres come together, 
merge and eventually disappear above the critical distance. As we demonstrate 
in this paper, the gravitational lensing signature of binary black holes 
is radically different depending on whether the photon spheres are present. 
Thus the separation between the black holes for fixed mass dictates the 
existence or otherwise of twin photon spheres and the gravitational lensing 
signature of the binary black holes. 

\section{Gravitational lensing by the Majumdar-Papapetrou di-hole}

In this section we describe the basics of the gravitational lensing formalism 
that we employ in our investigation of Majumdar-Papapetrou di-hole. We assume 
that the source of light and observer are located on the symmetry plane $z=0$ 
placed midway between the black holes. We restrict our attention to the 
light rays that are allowed to move on the plane $z=0$. This allows us 
to employ the formalism developed for the lensing of light on the equatorial 
plane of spherically symmetric spacetimes. We note that the central point with 
$\rho=0$ on the plane $z=0$, which is also the point midway between the two 
identical black holes, essentially acts as a gravitational lens in our 
investigation for all practical purposes.

\begin{figure}
\begin{center}
\includegraphics[width=0.65\textwidth]{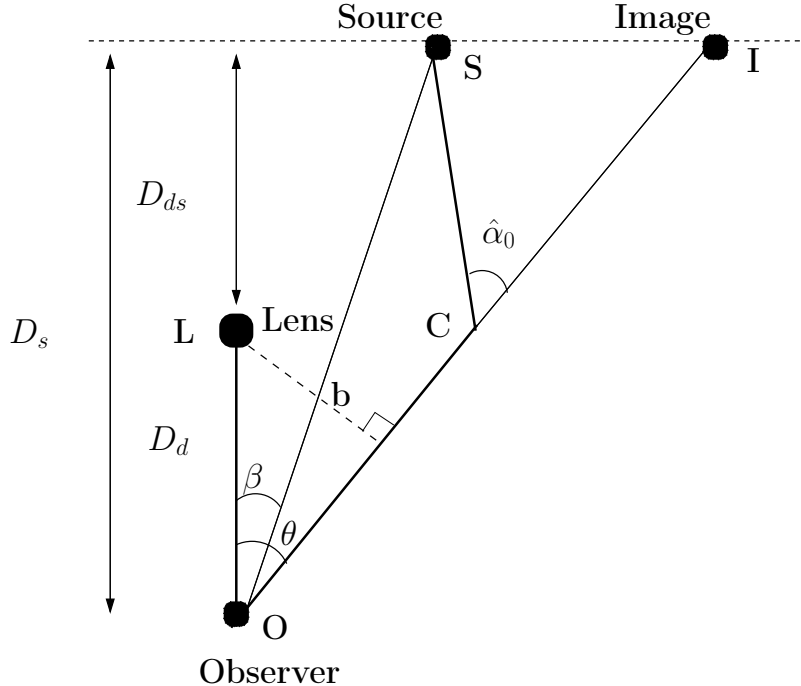}
\caption{\label{lens}
The lens diagram. $S$, $O$, $I$ and $L$ stand for the source, observer, image 
and lens respectively. $\hat{\alpha}_0$ is the deflection angle. $D_{ds}$, 
$D_d$ and $D_s$ are lens-image, lens-observer and source-observer distances 
respectively. Angle $\beta$ depicts the source location and $\theta$ depicts 
the image location. $b$ is the impact parameter.}
\end{center}
\end{figure}

We assume that both source and observer are located faraway from the central 
point and also from the two black holes in the asymptotic region which is 
approximately flat. Thus the light ray starts from infinity, falls toward the 
center as it gets bent, admits a turning point at $\rho=\rho_{0}$ and returns 
to infinity. The total amount of deflection suffered by the light ray 
in its journey is given by 
\begin{equation}
\hat{\alpha}(\rho_0)=2\int_{\rho_0}^{\infty}\frac{|V^{\phi}|}{|V^{\rho}|}
d\rho-\pi= 2\int_{\rho_{0}}^{\infty}\frac{1}{\rho} 
\frac{\sqrt{V_{eff}(\rho)}}{\sqrt{V_{eff}(\rho_0)-V_{eff}(\rho)}}d\rho
-\pi .
\label{defl}
\end{equation}
While obtaining the equation above from \eq{vel} we have used the fact that 
radial component of velocity is zero at the turning point, i.e., 
$V^{\rho}(\rho_0)=0$ and thus the impact parameter $b$ can be related 
to $\rho_0$ via 
\begin{equation}
 b=\frac{1}{\sqrt{V_{eff}(\rho_0)}}.
 \label{b1}
\end{equation}
Later in the paper we employ clever techniques proposed by Bozza 
\cite{bozza2} and develop their generalizations in order to get an 
approximate expression for deflection angle in various situations. 

We further assume that the source is located approximately behind the lens with 
respect to the observer. This allows us to use Virbhadra-Ellis lens equation 
\cite{virbhadra}. We note that there are other lens equations in the literature 
which also work equally well in this approximation such as 
Virbhadra-Narasimha-Chitre lens equation proposed earlier in \cite{narasimha} 
and so on.

The lens diagram is depicted in Fig.\ref{lens}. $L$ in the diagram is the lens 
which is central point of $z=0$ plane in this context. The observer is located 
at $O$. The line joining observer and lens is the optic axis. $S$ is the location 
of source which is almost exactly behind the lens with respect to observer. The 
source is located at an angle $\beta$ with respect to optic axis as seen by the 
observer. The light ray which initially travels along line 
$SC$ gets bent in the vicinity of the lens and arrives at the observer moving 
along $CO$. Thus to the observer the light seems to originate from $I$ 
which is the perceived image. The image is located at angle $\theta$ with 
respect to the optic axis. The deflection suffered by the light ray on its 
journey from source to the observer is $\hat{\alpha}_0$. The light can go 
around the lens multiple times. Hence the deflection angle can be very large. 
The distances between lens-image, lens-observer and source-observer are given by
$D_{ds}$, $D_{d}$ and $D_{s}$ respectively.  

The Virbhadra-Ellis lens equation allows us to relate the source location 
$\beta$, image location $\theta$ and the deflection angle $\hat{\alpha}_0$ and 
is given by
\begin{equation}
 \tan \beta=\tan \theta- \frac{D_{ds}}{D_{s}}\left(\tan \theta+ \tan 
 \left(\hat{\alpha}_0 - \theta\right)\right).
 \label{ve}
\end{equation}
Another relation that we can write down from the lens diagram is 
\begin{equation}
 \sin\theta=\frac{b}{D_d}.
 \label{b2}
\end{equation}
From Eqs.(\ref{defl}),(\ref{b1}),(\ref{ve}) we can write deflection angle 
$\hat{\alpha}_0$ in terms of $\theta$ and then use lens equation Eq.(\ref{ve}) 
to solve for the image locations $\theta$ for a given source location $\beta$. 
Solving lens equation is not so easy as it is a complicated transcendental 
equation. One has to often resort to numerical techniques. Analytical way 
of solving lens equation in certain situations was proposed by Bozza in 
\cite{bozza2} which we generalize and develop further in this paper. 
For this we use the fact that the source is almost exactly behind the source 
and thus angles $\beta$ and $\alpha$ are small and deviation of 
deflection angle from multiple of $2\pi$ is very small, i.e.,
\begin{equation}
 \hat{\alpha}_0=2\pi n + \delta \alpha_n ~~;~~ |\delta \alpha_n|<<1 . 
 \label{deflapp}
\end{equation}
This allows us to simplify lens equation Eq.(\ref{ve}) and Eq.(\ref{b2}) as 
\begin{equation}
 \beta=\theta - \frac{D_{ds}}{D_{s}}\delta \alpha_n ,
 \label{lensapp}
\end{equation}
and 
\begin{equation}
 \theta=\frac{b}{D_{d}}. 
 \label{bth}
\end{equation}
We will employ these equations later in the paper. 

The lens equation allows us to relate image location to the source location and 
set up a map from image plane to source plane. A radial caustic is admitted if 
this map is degenerate, i.e. 
\begin{equation}
 \frac{d\beta}{d\theta}=0.
\end{equation}
We now derive an expression for $\frac{d\beta}{d\theta}$. From Eq.(\ref{ve}), we
get 
\begin{equation}
\frac{d\beta}{d\theta}= \frac{\cos^2 \beta}{\cos^2 \theta}\ \left[ 
1-\frac{D_{ds}}{D_{s}} 
\left\{ 1+\frac{\cos^2\theta}{\cos^2(\hat{\alpha}_0 - \theta)} \left( 
\frac{d\hat{\alpha}_0}{d\rho_0} \frac{d\rho_0}{d\theta} -1  \right) \right\} 
\right] ,
\end{equation}
where $\frac{d\rho_0}{d\theta}$ and $\frac{d\hat{\alpha}_0}{d\rho_0}$ are as 
given below. From Eqs.(\ref{b1}),(\ref{b2}) we get
\begin{equation}
 \frac{d\rho_0}{d \theta}= -2\frac{V_{eff}(\rho_0)}{V^{'}_{eff}(\rho_0)} 
 \sqrt{D_{d}^2V_{eff}(\rho_0)-1},
 %\frac{\sqrt{D_{d}^2 - \rho_0^2 U(\rho_0)^4} }{U(\rho_0)\left(U(\rho_0)
%  +2\rho_0 U^{'}(\rho_0)  \right)},
\end{equation}
and from Eq.(\ref{defl}) after implementing few clever tricks we get 
\begin{equation}
 \frac{d\hat{\alpha}_0}{d\rho_0}= 
 2\int_{\rho_0}^{\infty} \left[ \frac{1}{\sqrt{V_{eff}(\rho_0) - 
V_{eff}(\rho)}} \frac{d}{d\rho} \left( \frac{\sqrt{V_{eff}(\rho)}}{\rho}  
\right)-\frac{1}{2}\frac{\sqrt{V_{eff}(\rho)} \left( 
V_{eff}^{'}(\rho_0)-V_{eff}^{'}(\rho) \right) }{ \rho \left( V_{eff}(\rho_0) 
-V_{eff}(\rho) \right)^{\frac{3}{2}} } \right]d\rho ,
\end{equation}
where prime denotes the derivative with respect to $\rho$. Radial caustic  
occurs when two images for a given source location that are separated from one another 
along the radial sense with respect to the lens coincide, has never been 
reported so far in the context of relativistic gravitational lensing. 
We will show that the radial caustic is admitted in the absence of the photon 
sphere, when the distance between two black holes is larger than a critical 
threshold value in Majumdar-Papapetrou di-hole spacetime.

Whenever required, for numerical computation we choose the following set of 
convenient parameters. We assume that the source and observer are 
equidistant from the lens. The mass of each of the black hole in binary is taken 
to be same as the mass of the galactic central supermassive black hole in Milky 
way which is $4.1 \times 10^{6} M_{\odot}$ and also we take distance of observer 
from lens to be same as the distance of earth from central supermassive black 
hole, i.e. $8.5 \text{kpc}$. The angular location of source is taken to be 
$\beta= 0.075 ~\text{radian}$.

Three cases $M>M_{*}$, $M=M_{*}$ and $M<M_{*}$ when two photon spheres are 
present, when a single degenerate photon sphere is present and when no photon 
spheres are present respectively require separate consideration and analysis and 
thus they are dealt with independently in the subsequent sections. 
Our aim is to analyze the structure of images and existence of caustic 
in the near-aligned configuration of source, lens and observer.

\begin{figure}
\begin{center}
\includegraphics[width=0.9\textwidth]{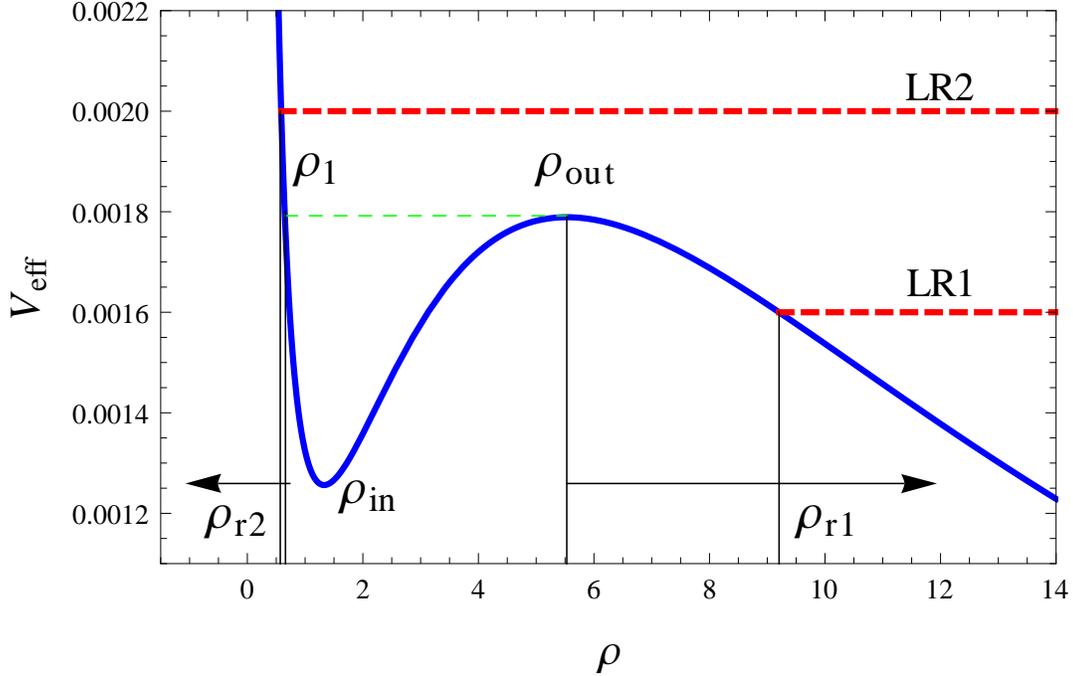}
\caption{\label{v1}
The effective potential $V_{eff}$ is plotted as a function of $\rho$ for 
$M=3>M_{*}$ denoted by blue curve. It admits a maximum at $\rho=\rho_{out}$ and 
a minimum at $\rho=\rho_{in}$. As shown in the figure effective potential rises 
again below the inner photon sphere for the smaller values of $\rho$. At 
$\rho=\rho_1$, we have $V_{eff}(\rho_1)=V_{\rho_{out}}$ i.e. the value of the 
effective potential is same as that at maximum. When 
$b>1/\sqrt{V_{eff}(\rho_{out})}$, light rays turn back from the region outside 
the outer photon sphere denoted by arrow which points towards right. Such a 
light ray depicted by $LR1$ which turns back from 
$\rho=\rho_{r1}>\rho_{out}$. Whereas the light rays with 
$b<1/\sqrt{V_{eff}(\rho_{out})}$ turn back from the region below 
$\rho=\rho_{1}$ denoted by arrow which points towards left. Such a light ray is 
denoted by $LR2$ which enters the outer photon sphere and turns back from 
$\rho=\rho_{r2}<\rho_1$.}
\end{center}
\end{figure}

\section{Two photon spheres}

In this section we focus our attention on the case $M>M_{*}$ when two photon 
spheres are present on the symmetry plane $z=0$ centered at $\rho=0$. The 
locations of the outer unstable photon sphere $\rho_{out}$ and inner 
stable photon sphere $\rho_{in}$ are given by Eqs.(\ref{rout}),(\ref{rin}). The 
effective potential admits maximum at $\rho=\rho_{out}$ and minimum at 
$\rho=\rho_{in}$ as shown in Fig.\ref{v1}. For lower values of $\rho$, 
the effective potential rises upwards below the inner photon sphere for the lower radii.
We note that we talk about the stability with respect to the radial perturbations 
restricting ourselves in $z=0$ plane. The situation can be different as far 
the stability in the two dimensional $(\rho-z)$ plane is concerned as analyzed in \cite{Dolan}.
$\rho=\rho_1$ is a point such that the effective potential at this location 
is same as the effective potential at maximum, i.e. $V_{eff}(\rho_1)=V_{eff}
(\rho_{out})$. The light rays for which $b>1/\sqrt{V_{eff}(\rho_{out})}$ turn 
back from $\rho_{0}>\rho_{out}$. Whereas the light rays for which 
$b<1/\sqrt{V_{eff}(\rho_{out})}$, enter outer photon sphere and admit turning 
point below $\rho=\rho_{1}$, i.e. $\rho_{0}<\rho_{1}$. This situation is quite 
different in case of the black holes where the effective potential does not rise 
again for lower values of radial coordinates and light rays which enter the 
photon sphere necessarily enter the event horizon. Thus new relativistic 
images are formed due to the lights rays which turn back in the region below 
outer photon sphere in dihole spacetime which are absent in black hole case.

\subsection{Images due to the light rays that turn back outside outer photon 
sphere.}

Initially we focus on the light rays that admit turning point at $\rho=\rho_0$ 
outside the outer photon sphere and calculate the location of 
the images formed. Further we assume that the deflection point 
is in fact very close to the photon sphere. This greatly simplifies the 
discussion and allows us to do calculations analytically. 

We find it convenient to introduce new radial coordinate $y$ following the 
discussion in \cite{bozza2} which is related to the old coordinate $\rho$ by 
\begin{equation}
 y=\frac{\left(\frac{1}{U(\rho)}-\frac{1}{U(\rho_0)}\right)}{\left(1-
 \frac{1}{U(\rho_0)}\right)}.
 \label{yco}
\end{equation}
$U$ is a monotonically decreasing function. Thus it follows that $y$ 
increases monotonically from $0$ to $1$ as $\rho$ varies from $\rho_0$ to 
$\infty$. It is also useful to invert the relation above and write $\rho$ in 
terms of $y$ as 
\begin{equation}
 \rho(y)=U^{-1}\left( \frac{U(\rho_0)}{U(\rho_0)+1-y} \right) ,
\end{equation}
where $U^{-1}$ is an inverse function of $U$.
We define a function $T(\rho_0,\rho)$ as 
\begin{equation}
 T(\rho_0,\rho) = \frac{d\rho}{dy}=-\frac{U(\rho)^2}{U^{'}(\rho)}\left(1-
 \frac{1}{U(\rho_0)}\right). 
\end{equation}

The deflection angle $\hat{\alpha}_0$ can be written as 
\begin{equation}
 \hat{\alpha}_0= I -\pi .
 \label{alphay}
\end{equation}
$I$ in the expression above is 
\begin{equation}
 I=\int_{\rho_0}^{\infty} \frac{2}{\rho} \frac{\sqrt{V_{eff}(\rho)}}{ 
 \sqrt{V_{eff}(\rho_0) - V_{eff}(\rho)}} d\rho = \int_{0}^{1} 
\frac{F(\rho_0,\rho(y))}{ \sqrt{V_{eff}(\rho_0) - V_{eff}(\rho(y))} } dy ,
 \label{int1}
\end{equation}
where $F(\rho_0,\rho(y))$ is a function given by 
\begin{equation}
 F(\rho_0,\rho(y))= \frac{2}{\rho} \sqrt{V_{eff}(\rho(y))} T(\rho_0,\rho(z)).
\end{equation}

The function $ F(\rho_0,\rho(y))$ is finite and well-behaved for all values of 
$y \in (0,1)$. Whereas $1/\sqrt{V_{eff}(\rho_0) - V_{eff}(\rho(y))}$ diverges 
at $y=0$. Taylor expanding $\left(V_{eff}(\rho_0) - V_{eff}(\rho(y))\right)$
around $y=0$, we get 
\begin{equation}
V_{eff}(\rho_0) - V_{eff}(\rho(y))= \alpha_{1}(\rho_0)y+\beta_{2}(\rho_0)y^2
+O(y^3) ,
\end{equation}
where $\alpha_{1}(\rho_0)$ and $\beta_1(\rho_0)$ are given by 
\begin{eqnarray}
 &&\beta_{1}(\rho_0)=-\frac{1}{2}\left( T^2(\rho_0,\rho=\rho_0) V_{eff}^{''}
 (\rho_0)+T(\rho_0,\rho=\rho_{0}) T^{'}(\rho_0,\rho=\rho_{0}) 
V_{eff}^{'}(\rho_0) \right)
 , \nonumber \\
 &&\alpha_{1}(\rho_0)=-F(\rho_0,\rho=\rho_0)V_{eff}^{'}(\rho_0) .
\end{eqnarray}
When turning point $\rho=\rho_0$ is away from the outer photon sphere 
$\rho=\rho_{out}$, $\alpha_{1}(\rho_0)$ is a non-zero positive finite number. 
Thus the integral Eq(\ref{int1}) and the deflection suffered by the light ray is 
finite. 

The situation is quite different when the turning point is close to the outer 
photon sphere. We expand $\alpha_{1}(\rho_0)$ and $\beta_{1}(\rho_0)$ around 
$\rho_0=\rho_{out}$ and get 
\begin{eqnarray}
&&\beta_{1}(\rho_0)=-\frac{1}{2} T^2(\rho_{out},\rho=\rho_{out}) V_{eff}^{''}
(\rho_{out}) + O(\rho_0-\rho_{out}) , \nonumber \\
&&\alpha_{1}(\rho_0)= \frac{2 \beta_{1}(\rho_{out})}{T(\rho_{out},
\rho=\rho_{out})} (\rho_{0}-\rho_{out}) +O((\rho_0-\rho_{out})^2) .
\label{ab1}
\end{eqnarray}
To the leading order $\alpha_{1}(\rho_0)$ is vanishingly small and thus the 
integral Eq(\ref{int1}) and the deflection angle show divergence. We isolate 
the divergence piece in the integral Eq(\ref{int1}) as 
\begin{equation}
 I_{1D}(\rho_0) =F(\rho_{out},\rho=\rho_{out}) \int_{0}^{1}\frac{1}{\sqrt{
 \alpha_{1}(\rho_0)y+\beta_{2}(\rho_0)y^2}}dy 
 = -A_{1} \log\left(\frac{\rho_0}{\rho_{out}}-1\right)+\tilde{B}_1+O(\rho_0-\rho_{out}) ,
\end{equation}
where 
\begin{eqnarray}
&& A_1=\frac{F(\rho_{out},\rho=\rho_{out})}{\sqrt{\beta_1(\rho_{out})}} , 
\nonumber \\
&& \tilde{B}_1= \frac{F(\rho_{out},\rho=\rho_{out})}{\sqrt{\beta_1(\rho_{out})}}
\log \left(\frac{2T(\rho_{out},\rho=\rho_{out})}{\rho_{out}} \right) ,
\end{eqnarray}
and the regular piece in the integral can be written as 
\begin{equation}
I_{1R} (\rho_0) =\int_{0}^{1}\left( \frac{F(\rho_0,\rho(y))}{\sqrt{V_{eff}
(\rho_0)-V_{eff}(\rho(y))}} - \frac{F(\rho_{out},\rho=\rho_{out})}{  
\sqrt{\alpha_{1}(\rho_0)y+\beta_{2}(\rho_0)y^2 }}  \right) dy = I_{1R} 
(\rho_{out}) + O(\rho_0 -\rho_{out}) .
\label{i1r}
\end{equation}
Combining divergent part $I_{1D}(\rho_0)$ and regular parts of the integral 
$I_{1R}(\rho_0)$, we can write down the deflection angle 
$\hat{\alpha}_0$ as 
\begin{equation}
 \hat{\alpha}_0=-A_{1} \log\left[B_1\left(\frac{\rho_0}{\rho_{out}}-1\right)
 \right]-\pi+O(\rho_0-\rho_{out}),
 \label{defl1}
\end{equation}
where $B_1$ is related to $\tilde{B}_1$, $A_1$ and $I_{1R}(\rho_{out})$ by  
\begin{equation}
B_1= \text{exp}\left(-\frac{\tilde{B}_1 + I_{1R}(\rho_{out})}{A_1}\right).
\end{equation}
Thus the deflection angle shows logarithmic divergence as the reflection point 
approaches the outer photon sphere. 

The impact parameter $b$ upon expansion around the outer photon sphere can be 
written as 
\begin{equation}
 b=C_1+D_1 \left( \frac{\rho_0}{\rho_{out}}-1 \right)^2+O\left( (\rho_0-\rho_{out})^3 \right),
 \label{bcd1}
\end{equation}
where $C1$ and $D_1$ are given by 
\begin{eqnarray}
 &&C_1=\frac{1}{\sqrt{V_{eff}(\rho_{out})}} , \nonumber \\
 &&D_1=-\frac{1}{2}   \frac{V_{eff}^{''}(\rho_{out})}{V_{eff}^{\frac{3}{2}}
 (\rho_{out}) } \rho_{out}^2 .
 \label{cd1}
\end{eqnarray}
Using Eqs.(\ref{bth}),(\ref{defl1}),(\ref{bcd1}) we can relate $\theta$ to 
the deflection angle as 
\begin{equation}
 \theta=\frac{C_1}{D_d}+\frac{D_1}{D_d}\frac{1}{B_1^2}\text{exp}\left(-
 \frac{2}{A_1}\left( \hat{\alpha}_0+\pi \right)\right) ,
 \label{thal}
\end{equation}
and using Eqs.(\ref{deflapp}),(\ref{lensapp}),(\ref{thal}) we can compute 
locations of images $\theta_{1,n}$ for a given source location $\beta$ in the 
near-aligned configuration, which are given by 
\begin{equation}
 \theta_{1,n}=\frac{C_1}{D_d}+  \frac{D_1}{D_d}\frac{1}{B_1^2}\text{exp}\left(
 -\frac{2}{A_1}(2n+1)\pi \right)
 \left(1+\frac{2}{A_1}\frac{D_{s}}{D_{ds}}\beta\right) .
\end{equation}
Here $n$ stands for the number of times light ray goes around the lens during 
its journey from source to observer. It turns out that all the images lie 
beyond a certain critical angle $\bar{\theta}_1$ given by 
\begin{equation}
 \bar{\theta}_1=\frac{C_1}{D_d}.
 \label{thcr1}
\end{equation}
As $n$ increases images get closer and closer to the critical angle 
$\bar{\theta}_1$ and asymptotically approach it from right. So far we dealt 
with the light rays which go around the lens in the clockwise sense. For the 
light rays which go around the lens in counter-clockwise sense the 
images occur on the opposite side of the optic axis and their locations 
$\theta^{'}_{1,n}$ are given by 
\begin{equation}
 \theta_{1,n}^{'}= -\frac{C_1}{D_d}+  \frac{D_1}{D_d}\frac{1}{B_1^2}\text{exp}
 \left(-\frac{2}{A_1}(2n+1)\pi \right)
 \left(-1+\frac{2}{A_1}\frac{D_{s}}{D_{ds}}\beta\right) .
\end{equation}

The pattern of the images formed due to the single black hole is qualitatively 
similar to the images formed in case of the di-hole due to the light rays that 
are reflected back outside the outer photon sphere. In case of the black hole 
there is a dark region below the critical radius as there are no images formed 
in this region. As we show in this paper that is not the case for di-hole 
as the light rays which turn back from the region inside photon sphere form 
new images filling in the void. 

\subsection{Images formed due to the light rays that turn back inside the 
photon sphere.}

\begin{figure}
\begin{center}
\includegraphics[width=0.8\textwidth]{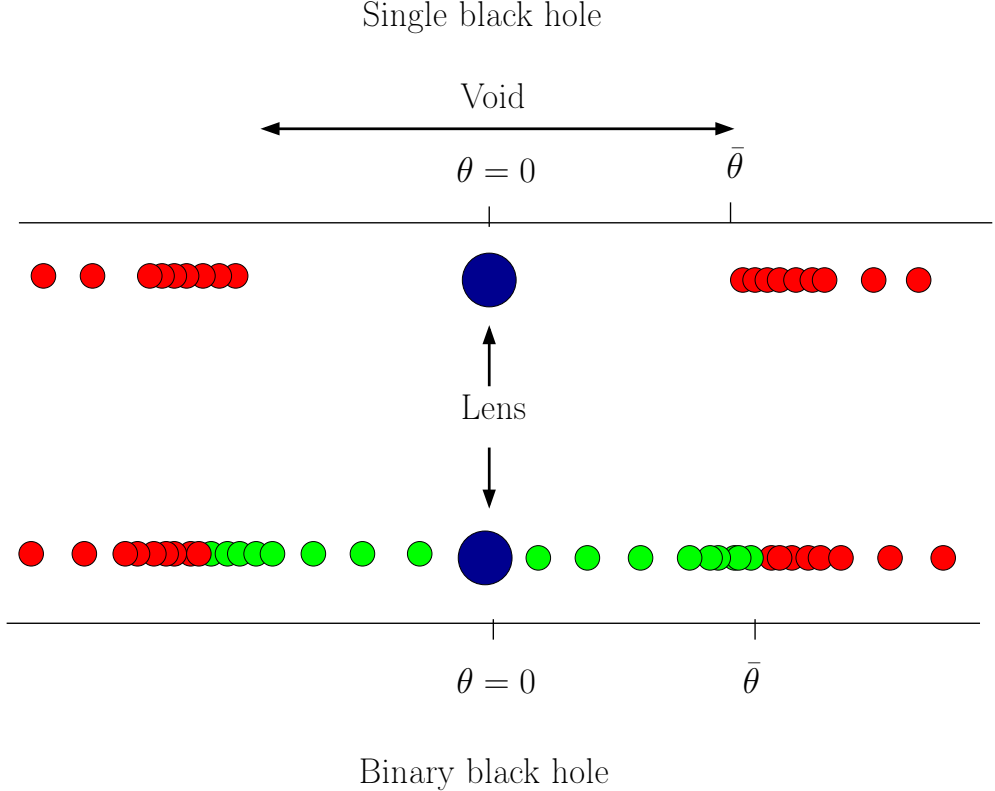}
\caption{\label{bbhimg}
Pattern of images in the case of single black hole (upper panel) and binary 
black hole (lower panel). In case of the single black hole images are formed 
due to the light rays that turn back outside the photon sphere and lie beyond 
critical angle $\bar{\theta}$ with respect to the lens. There is a dark region 
below the critical angle denoted as "Void" in the figure where no images are 
formed. This region is also known as "shadow".
In case of the binary black hole there are two photon spheres when 
$M>M_{*}$ and light rays which enter photon sphere can turn back and hence we 
get new images. Red dots in the lower panel are images formed due to the light 
rays which turn back outside the outer photon sphere and lie above the 
critical angle $\bar{\theta}$ with respect to lens. Whereas the green dots represent the images 
formed due to the light rays which admit turning point inside the photon sphere, 
all of which lie below the critical angle $\bar{\theta}$. Thus a new set of 
relativistic images fills in the void present in case of the single black hole 
and illuminate the dark region.}
\end{center}
\end{figure}

We now compute the pattern of images formed due to the light rays that enter the
outer photon sphere and turn back at the radial coordinate $\rho_0<\rho_1$ and 
reach infinity again. As mentioned earlier $\rho_1$ is a location such that 
\begin{equation}
 V_{eff}(\rho_1)=V_{eff}(\rho_{out}).
 \label{rhoeq}
\end{equation}
We use the coordinate $y$ introduced earlier Eq.(\ref{yco}). The deflection 
angle can be written as before in Eqs.(\ref{alphay}),(\ref{int1})
\begin{equation}
  \hat{\alpha}_0= I -\pi = \int_{0}^{1} \frac{F(\rho_0,\rho(y))}{ 
  \sqrt{V_{eff}(\rho_0) - V_{eff}(\rho(y))} } dy -\pi .
\end{equation}
$F(\rho_0,\rho(y))$ is finite and well-behaved everywhere . On the other hand 
$1/\sqrt{V_{eff}(\rho_0) - V_{eff}(\rho(y))}$ diverges when $y=0$ i.e. at 
$\rho=\rho_0$ and at $\rho=\rho_{out}$ i.e. $y=y_{out}$ when $\rho_0$ is close 
to $\rho_1$. Here $y_{out}$
is given by $y_{out}=(1/U(\rho_{out})-1/U(\rho_0))/(1-1/U(\rho_0))$.

We first focus on $\rho=\rho_0$ and expand $V_{eff}(\rho_0) - V_{eff}(\rho(y))$ 
around $y=0$. To the leading order we obtain
\begin{equation}
 V_{eff}(\rho_0) - V_{eff}(\rho(y))=\alpha_2(\rho_0)y+O(y^2),
\end{equation}
where $\alpha_2(\rho_0)$ is given by
\begin{equation}
 \alpha_2(\rho_0)=-T^{'}(\rho_0,\rho=\rho_0) V_{eff}^{'}(\rho_0).
\end{equation}
Since $ \alpha_2(\rho_0)$ is a finite, positive non-zero number, divergent 
behavior of integrand can be tamed when we compute the integral $I$ and it does 
not result in the divergent behavior of deflection angle. 

We now focus on $\rho=\rho_{out}$ and expand $V_{eff}(\rho_0) - 
V_{eff}(\rho(y))$ around $y=y_{out}$. We get 
\begin{equation}
 V_{eff}(\rho_0) - V_{eff}(\rho(y))= \alpha_{3}(\rho_{out})+
 \beta_{3}(\rho_{out})(y-y_{out})^2+O\left( (y-y_{out})^3 \right) ,
\end{equation}
where $\alpha_{3}(\rho_{out})$ and $\beta_{3}(\rho_{out})$ are given by 
\begin{eqnarray} 
&&\alpha_{3}(\rho_{out})=V_{eff}(\rho_0)-V_{eff}(\rho_{out})=-V_{eff}^{'}
(\rho_1)(\rho_1-\rho_0)+O\left( (\rho_1-\rho_0)^2 \right) , \nonumber \\
 && \beta_{3}(\rho_{out})=-\frac{1}{2}T^{2}(\rho_1,\rho=\rho_{out}) V_{eff}^{''}
 (\rho_{out})+O\left( \rho_0-\rho_1 \right).
\end{eqnarray}
We used Eq.(\ref{rhoeq}) to derive the expression above. When $\rho_0$ is very 
close to $\rho_1$, integral $I$ is divergent and so is the deflection 
angle $\hat{\alpha}_0$.

We identify the divergent piece in the integral $I_{2D}$ as 
\begin{equation}
 I_{2D}= F(\rho_1,\rho=\rho_{out}) \int_{0}^{1} \frac{1}{\sqrt{\alpha_3+\beta_3
 (y-y_{out})^2}} dy = -A_2\log\left(1-\frac{\rho_0}{\rho_1}\right)+\tilde{B}_2+O(\rho_0 -\rho_1) ,
\end{equation}
where 
\begin{eqnarray}
 && A_2=\frac{F(\rho_1,\rho=\rho_{out})}{\sqrt{\beta_3}}  , \nonumber \\
 &&\tilde{B}_2= \frac{F(\rho_1,\rho=\rho_{out})}{\sqrt{\beta_3}} 
 \log\left( \frac{4 y_{out}(1-y_{out})\beta_3}{-V_{eff}^{'}(\rho_1)\rho_1} 
\right) ,
\end{eqnarray}
and the regular piece in the integral $I_{2R}$ can be written as 
\begin{equation}
I_{2R} (\rho_0) =\int_{0}^{1}\left( \frac{F(\rho_0,\rho(y))}{\sqrt{V_{eff}
(\rho_0)-V_{eff}(\rho(y))}} - \frac{F(\rho_{1},\rho=\rho_{out})}{  
\sqrt{\alpha_{3} + \beta_{3}(y-y_{out})^2 }}  \right) dy = I_{2R} (\rho_{1}) + 
O(\rho_1 -\rho_{0}) .
\label{i2r}
\end{equation}
While writing the equations above we set $\rho_0=\rho_1$ in the expression for
$y_{out}$ since $y_{out}(\rho_0)=y_{out}(\rho_1)+O(\rho_1-\rho_0)$ and hence 
the error involved is at higher order. Combining the divergent piece $I_{2D}$ 
and regular piece $I_{2R}$ in the integral we can write the deflection angle 
$\hat{\alpha}_0$ as 
\begin{equation}
\hat{\alpha}_0= -A_2 \log \left( B_2 \left(1-\frac{\rho_0}{\rho_1}\right)\right)
-\pi + O(\rho_0 -\rho_1),
\label{defl3}
\end{equation}
where 
\begin{equation}
 B_2= \text{exp} \left( -\frac{\tilde{B}_2+I_{2R}(\rho_1)}{A_2} \right).
\end{equation}
The deflection angle again shows the logarithmic divergence as the reflection 
point $\rho_0$ approaches $\rho_1$. We now write the impact parameter $b$, 
expanding it around $\rho=\rho_1$. We get 
\begin{equation}
 b=C_2-D_2\left(1-\frac{\rho_1}{\rho_0}\right) + O\left(\left(\rho_0-
 \rho_1\right)^2 \right),
 \label{bcd3}
\end{equation}
where $C_2$ and $D_2$ are given by 
\begin{eqnarray}
 &&C_2=\frac{1}{\sqrt{V_{eff}(\rho_1)}} \nonumber \\
 && D_2=-\frac{1}{2}\frac{V_{eff}^{'}(\rho_1)}{V_{eff}^{\frac{3}{2}}(\rho_1)}
 \rho_1 .
 \label{cd2}
\end{eqnarray}

Using Eqs.(\ref{bth}),(\ref{defl3}),(\ref{bcd3}) we can relate $\theta$ to the 
deflection angle $\hat{\alpha}_0$ as
\begin{equation}
 \theta=\frac{C_2}{D_d}+\frac{D_2}{D_d}\frac{1}{B_2}\text{exp}\left(-
 \frac{1}{A_2}\left( \hat{\alpha}_0+\pi \right)\right) ,
 \label{thag}
\end{equation}
and using Eqs.(\ref{deflapp}),(\ref{lensapp}),(\ref{thag}) we can compute 
locations of images $\theta_{2,n}$ for a given source location $\beta$ in the 
near-aligned configuration, which are given by 
\begin{equation}
 \theta_{2,n}=\frac{C_2}{D_d} - \frac{D_2}{D_d}\frac{1}{B_2^2}\text{exp}
 \left(-\frac{1}{A_2}(2n+1)\pi \right)
 \left(1+\frac{1}{A_2}\frac{D_{s}}{D_{ds}}\beta\right) .
\end{equation}
All the images lie below the critical angle $\bar{\theta}_2$ which is given by 
\begin{equation}
 \bar{\theta}_{2}=\frac{C_2}{D_d}.
 \label{thcr2}
\end{equation}
As we increase $n$, images get closer and closer to critical angle and 
asymptotically approach it from left. From 
Eqs.(\ref{cd1}),(\ref{thcr1}),(\ref{rhoeq}),(\ref{cd2}),(\ref{thcr2}), we see 
that 
\begin{equation}
 \bar{\theta}_1=\bar{\theta}_2.
\end{equation}
This implies that the second set of images formed due to the light rays which 
enter photon sphere and turn back fill in what would have been a void and dark 
region in case of the single black hole at an angle $\theta$ less that the critical angle towards optic axis.

Here $\theta_{2,n}$ denote the location of images formed on the right side of 
the optic axis due to the light rays that move around the lens in the clockwise 
sense. The location of images formed on the left side of the optic axis due to 
the light rays which move in anti-clockwise sense are given by 
\begin{equation}
 \theta^{'}_{2,n}=-\frac{C_2}{D_d} + \frac{D_2}{D_d}\frac{1}{B_2^2}\text{exp}
 \left(-\frac{1}{A_2}(2n+1)\pi \right)
 \left(1 - \frac{1}{A_2}\frac{D_{s}}{D_{ds}}\beta\right) .
\end{equation}

We note that $I_{1R}(\rho_{out})$ and $I_{2R}(\rho_1)$ must be computed 
numerically since the integrals Eqs.(\ref{i1r}),(\ref{i2r}) are difficult to 
evaluate analytically. It might also be necessary to compute $\rho_1$ 
numerically since one needs to solve transcendental equation 
$V_{eff}(\rho_1)=V_{eff}(\rho_{out})$. All other calculations presented above 
can be carried out analytically. 

The pattern of images formed is shown in Fig.(\ref{bbhimg}). The images on the 
right side of the optic axis i.e. lens are associated with the light rays that 
move clockwise and images on the left side are associated with the light rays 
which move in the counter-clockwise sense. In case of a single black hole light 
rays which enter the photon sphere are doomed to enter the black hole. Thus 
images are formed due to the light rays that turn back outside the photon 
sphere. All images occur above the critical angle $\bar{\theta}$ and infinitely 
many images are clubbed together just above the critical angle as shown 
in the figure. No images are formed below the critical angle towards the 
lens resulting in the dark void. In case of the binary black hole due to the 
presence of inner stable photon sphere inside the outer unstable photon sphere 
effective potential turns upwards again and light rays which enter the outer 
photon sphere can turn back giving rise to another set of new relativistic 
images. The new images lie below the critical angle $\bar{\theta}$ as shown in 
the figure. Infinitely many images clubbed just below $\bar{\theta}$. Thus the 
region which would have  been dark void is filled up with images and turns 
bright again. Thus the pattern of images is qualitatively different in the 
binary black holes spacetime.

\begin{figure}
\begin{center}
\includegraphics[width=0.8\textwidth]{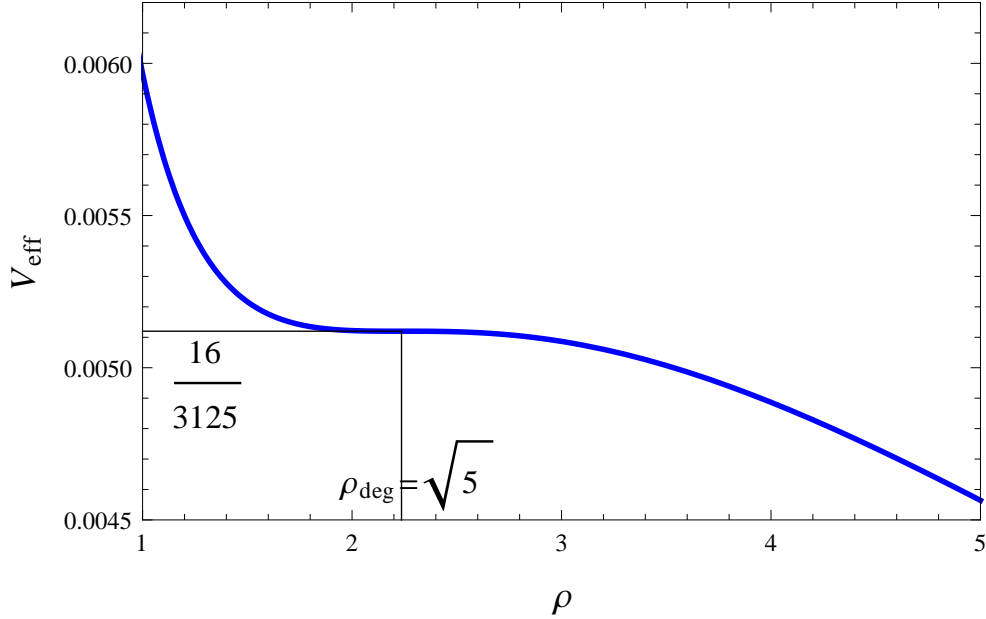}
\caption{\label{vdeg}
Effective potential $V_{eff}$ is plotted against $\rho$ for the critical value 
of the parameter $M=M_{*}=\sqrt{\frac{27}{8}}$. The two photon spheres coincide 
and we have a single photon sphere located at $\rho=\rho_{deg}=\sqrt{5}$. The 
value of the effective potential at the location of the degenerate 
photon sphere is $V_{eff}(\rho_{deg})=16/3125$ as shown in the figure. Both 
first as well as second derivatives are zero at the location of photon sphere, 
i.e., $V_{eff}^{'}(\rho_{deg})=V_{eff}^{''}(\rho_{def})=0$.}
\end{center}
\end{figure}

\section{A single degenerate photon sphere}

We now deal with the case where the parameter $M$ assumes the critical value 
$M=M_{*}=\sqrt{27/8}$. The two photon spheres we encountered in the earlier 
section when $M>M_{*}$ now coincide and we have a single degenerate photon 
sphere present at $\rho=\rho_{deg}=\sqrt{5}$, where both the first and second 
derivative of the effective potential vanish, i.e., 
$V_{eff}^{'}(\rho_{deg})=V_{eff}^{''}(\rho_{def})=0$. The value of the effective 
potential at the location of the degenerate photon sphere is given by 
$V_{eff}(\rho_{deg})=16/3125$. The light rays admit turning point at 
$\rho>\rho_{deg}$ when the impact parameter satisfies 
$b>1/\sqrt{V_{eff}(\rho_{deg})}$ and at $\rho<\rho_{deg}$ when the impact 
parameter is $b<1/\sqrt{V_{eff}(\rho_{deg})}$. We present here the unified 
treatment of the light rays that turn back both above and below the photon 
sphere. Such a treatment was not possible for $M>M_{*}$ since the light rays 
admitted turning points either at $\rho>\rho_{out}$ or $\rho <\rho_{1}$ where 
$\rho_{out}$ and $\rho_1$ were distinct points with qualitatively 
different behavior of effective potential $V_{eff}$.

Once again we use the new radial coordinate $y$ with the finite range 
introduced in Eq.(\ref{yco}). The deflection angle can be written as 
\begin{equation}
 \hat{\alpha}_0= I - \pi = \int_{0}^{1} \frac{F(\rho, \rho(y))}{\sqrt{ 
V_{eff}(\rho_0) -V_{eff}(\rho(y)) }}dy - \pi . 
\end{equation}
We request readers to refer to section IV-A for the definition of various
quantities that we use here. The function $F(\rho, \rho(y))$ is finite and 
well-behaved everywhere as earlier whereas $1/\sqrt{ V_{eff}(\rho_0) 
-V_{eff}(\rho(y)) }$ is divergent at $\rho=\rho_{0}$, i.e., at $y=0$. 

We expand $V_{eff}(\rho_0) -V_{eff}(\rho(y))$ around $y=0$ as 
\begin{equation}
 V_{eff}(\rho_0) -V_{eff}(\rho(y))= \alpha_4 (\rho_0) y + \beta_{4}(\rho_0) y^2 
 + \gamma_4 (\rho_0) y^3 + O\left(y^4\right),
\end{equation}
where 
\begin{eqnarray}
\alpha_4(\rho_0) = &&- T(\rho_0,\rho=\rho_0) V_{eff}^{'}(\rho_0)  ,  \\
 \beta_4(\rho_0) =&& -\frac{1}{2} \left(   T(\rho_0,\rho=\rho_0) T^{'}(\rho_0,
 \rho=\rho_0) V_{eff}^{'}(\rho_0) +
T^2(\rho_0,\rho=\rho_0)  V_{eff}^{''}(\rho_0) \right) , \nonumber \\
 \gamma_4 (\rho_0) = && -\frac{1}{6}  \left(   T(\rho_0,\rho=\rho_0) 
T^{'2}(\rho_0,
\rho=\rho_0) V_{eff}^{'}(\rho_0) +T^2(\rho_0,\rho=\rho_0) 
T^{''}(\rho_0,\rho=\rho_0) V_{eff}^{'}(\rho_0) \right) \nonumber \\
     && -\frac{1}{6} \left( 3  T^2(\rho_0,\rho=\rho_0) T^{'}(\rho_0,
     \rho=\rho_0)  V_{eff}^{''}(\rho_0)  +   T^{2}(\rho_0,\rho=\rho_0) 
V_{eff}^{'''}(\rho_0) \right)  
\end{eqnarray} 

If $\rho_0$ is away from the photon sphere radius $\rho=\rho_{deg}$ then 
$\alpha_4(\rho_0)$ is finite and the integral $I$ and deflection angle 
$\hat{\alpha}_0$ would be finite. If $\rho_0$ is reasonably close to 
$\rho=\rho_{deg}$ then $\alpha_4(\rho_0)$ is small but the $\beta_4(\rho_0)$ is 
finite when $V_{eff}^{'}(\rho_0)$ is close to zero, however 
$V_{eff}^{''}(\rho_0)$ is finite. In this intermediate region the integral $I$ 
and the deflection angle $\hat{\alpha}_0$ would diverge logarithmically. 
In this case the calculations would be identical to that in Section IV-A. 

When $\rho_0$ is sufficiently close to $\rho=\rho_{deg}$ both 
$V_{eff}^{'}(\rho_0)$
and $V_{eff}^{''}(\rho_0)$ take values 
which are close to zero. Expanding around $\rho_0=\rho_{deg}$, we get 
\begin{eqnarray}
&&\gamma_4(\rho_0)=-\frac{1}{6} T^{3}(\rho_{deg},\rho=\rho_{deg}) 
V_{eff}^{'''}(\rho_{deg})
+ O \left( \rho_0 - \rho_{deg}\right) , \nonumber \\
&&\alpha_4(\rho_0)= \frac{3}{T^2(\rho_{deg},\rho=\rho_{deg})} 
\gamma_4(\rho_{deg}) 
(\rho_0 -\rho_{deg})^2 + O\left(  \left(\rho_0 -\rho_{deg}\right)^3\right) , 
\nonumber \\
&&\beta_4(\rho_0)= \frac{3}{T(\rho_{deg},\rho=\rho_{deg})} \gamma_4(\rho_{deg}) 
(\rho_0 -\rho_{deg}) + O\left(  \left(\rho_0 -\rho_{deg}\right)^2\right) .
 \end{eqnarray}
Since $\alpha_4(\rho_0)$ and $\beta_4(\rho_0)$ tend to zero near 
$\rho_0=\rho_{deg}$, 
the integral $I$ and the deflection angle 
$\hat{\alpha}_0$ show divergence. We extract the divergent part in the integral 
$I$ as 
\begin{equation}
 I_{3D}=F(\rho_{deg}, \rho=\rho_{deg}) \int_{0}^{1} dy \frac{1}{\sqrt{ \alpha_4
 (\rho_0) y + \beta_{4}(\rho_0) y^2 + \gamma_4 (\rho_0) y^3 }}  ,
\end{equation}
which can be written as 
\begin{equation}
 I_{3D}=\frac{A_{3,\pm}}{\sqrt{ | 1-\frac{\rho_0}{\rho_{deg}} |  }} + \tilde{B}_{3D} +O(\rho_0 \rho_{deg}),
\end{equation}
where $A_{3,\pm}$ and $\tilde{B}_{3D}$ are given by 
\begin{eqnarray}
 && A_{3,\pm}= \frac{2 F(\rho_{deg},\rho=\rho_{deg})}{\gamma_4(\rho_{deg})} 
 \sqrt{T(\rho_{deg},\rho=\rho_{deg})} \mathcal{I}_{\pm} , \nonumber \\
 &&  \tilde{B}_{3D}= - \frac{2 F(\rho_{deg},\rho=\rho_{deg})}{\gamma_4(\rho_{deg})} ,
\end{eqnarray}
with $\mathcal{I}_{\pm}$ as 
\begin{equation}
 \mathcal{I}_{\pm}=\int_{0}^{\infty} d\xi \frac{1}{\sqrt{3 \pm 3 \xi^2+ \xi^{4}} }.
\end{equation}
The regular part of the integral $I$ is given by 
\begin{eqnarray}
 I_{3R}(\rho_0)=&& \int_{0}^{1} dy \left( 
\frac{F(\rho_0,\rho(y))}{\sqrt{V_{eff}(\rho_0) 
- V_{eff}(\rho(y))  }}  - \frac{F(\rho_{deg}, \rho=\rho_{deg} )}{ \sqrt{ 
\alpha_4 (\rho_0) y + \beta_{4}(\rho_0) y^2 + \gamma_4 (\rho_0) y^3 } }   
\right) \nonumber \\
=&& I_{3R}(\rho_{deg})+ O\left( \rho_0-\rho_{deg} \right) .
\end{eqnarray}
Combining the divergent and convergent parts of the integral we can write the 
deflection angle as 
\begin{equation}
 \hat{\alpha}_0=\frac{A_{3,\pm}}{\sqrt{ | 1-\frac{\rho_0}{\rho_{deg}} |  }} + B_3 - \pi +O(\rho_0 -\rho_{deg}),
 \label{defl4}
\end{equation}
where $B_3$ is given by 
\begin{equation}
B_3=\tilde{B}_{3D} + I_{3R}(\rho_{deg}). 
\end{equation}
The deflection angle shows the power law divergence unlike in the section IV 
where it diverged logarithmically. 

Expanding the impact parameter $b$ around $\rho_0=\rho_{deg}$ we get 
\begin{equation}
 b(\rho_0)= C_3 + D_{3,\pm} |1-\frac{\rho_0}{\rho_{deg}}|^3 + O\left( (\rho_0 -\rho_{deg})^4 \right) ,
\end{equation}
where 
\begin{eqnarray}
 && C_3= \frac{1}{V_{eff}(\rho_{deg})} , \nonumber \\
 && D_{3,\pm}= \mp \frac{1}{12} \frac{V_{eff}^{'''}(\rho_{deg})}{ 
V_{eff}^{\frac{3}{2}}(\rho_{deg})} 
\rho_{deg}^3 | 1-\frac{\rho_0}{\rho_{deg}}|^3 ,
\label{bcd4}
 \end{eqnarray}
where $\pm$ stands for the light rays that turn back at $\rho_0 > \rho_{deg}$ 
and $\rho_0 < \rho_{deg}$ respectively.

Using Eqs.(\ref{bth}),(\ref{defl4}),(\ref{bcd4}) we can relate $\theta$ to the 
deflection angle $\hat{\alpha}_0$ as
\begin{equation}
 \theta= \frac{C_3}{D_d}+ \frac{D_{3,\pm} A_{3,\pm}^3}{\left(  \hat{\alpha}_0 
 + \pi - B_3\right)^3}.
 \label{thag2}
\end{equation}
Using Eqs.(\ref{deflapp}),(\ref{lensapp}),(\ref{thag2}) we can compute 
locations 
of images $\theta_{3,n}$ for a given source 
location $\beta$ in the near-aligned configuration, which are given by 
\begin{equation}
 \theta_{3,n}=\frac{C_3}{D_d}+  \frac{D_{3,\pm} A_{3,\pm}^3}{\left(  (2n+1) 
 \pi - B_3\right)^3} \left( 1+ \frac{3}{\left(  (2n+1) \pi - B_3\right)} 
\frac{D_s}{D_{ds}} \beta \right) .
\end{equation}

\begin{figure}
\begin{center}
\includegraphics[width=0.8\textwidth]{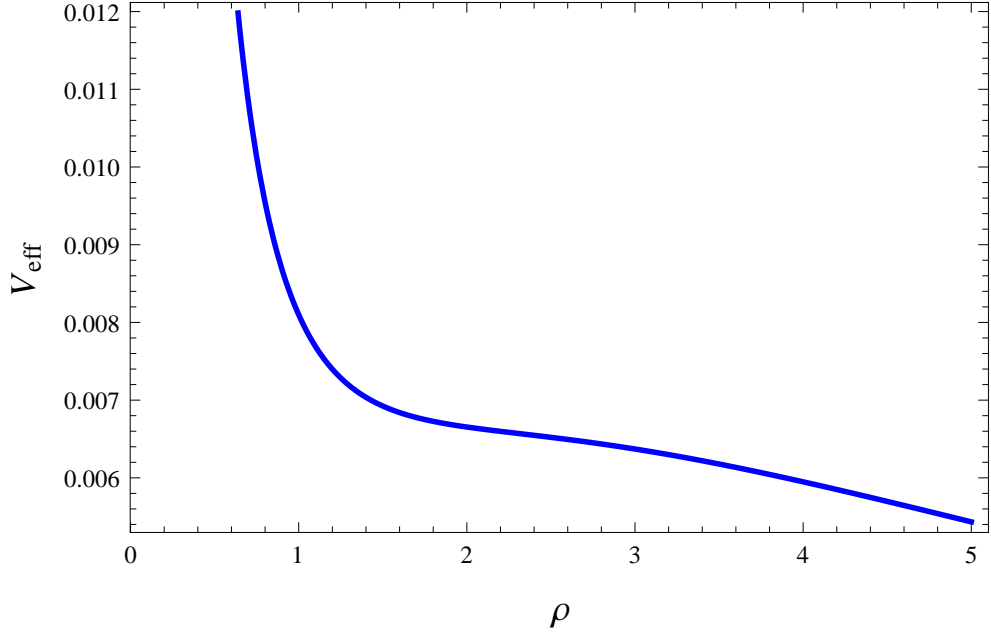}
\caption{\label{nph}
Effective potential $V_{eff}$ is plotted against $\rho$ for the value of the 
parameter $M$ below the criticality, i.e., $M<M_{*}$. The effective potential 
is monotonic function and does not admit any extremum. Thus photon sphere is 
absent.}
\end{center}
\end{figure}

For the light rays that turn back outside the photon sphere, i.e., $\rho_0 
>\rho_{deg}$, since $D_{3,+}>0$, the images are located above the critical 
angle $\bar{\theta}_3$ which is given by 
\begin{equation}
 \bar{\theta}_3=\frac{C_3}{D_d}.
\end{equation}
As $n$ increases, images get closer and closer to $\bar{\theta}_3$ from the 
right side. On the other hand for the light rays that get reflected back below 
the photon sphere, i.e., $\rho_0< \rho_{deg}$, we have $D_{3,-}<0$ and the 
images are located below $ \bar{\theta}_3$. The images get closer and tend 
towards  $\bar{\theta}_3$ from left. The pattern of images formed is 
quite similar to the one depicted in the lower panel of Fig.\ref{bbhimg}. The 
$\theta_{3,n}$ are locations of the images formed due to the light rays that 
move around the lens in the clockwise sense. The location of the images 
$\theta^{'}_{3,n}$ formed due to the light rays that move in the 
counter-clockwise sense are given by 
\begin{equation}
 \theta^{'}_{3,n}= - \frac{C_3}{D_d}+  \frac{D_{3,\pm} A_{3,\pm}^3}{\left(  
 (2n+1) \pi - B_3\right)^3} \left( -1+ \frac{3}{\left(  (2n+1) \pi - 
B_3\right)} \frac{D_s}{D_{ds}} \beta \right) .
\end{equation}
The analysis presented in this section is analytical except for the calculation 
of $I_{3R}(\rho_{deg})$ which is carried out numerically. In next section we 
discuss the case where $M<M_{*}$ when the black holes are far apart and no 
photon spheres are present.

\section{No photon sphere}

\begin{figure}
\begin{center}
\includegraphics[width=1.0\textwidth]{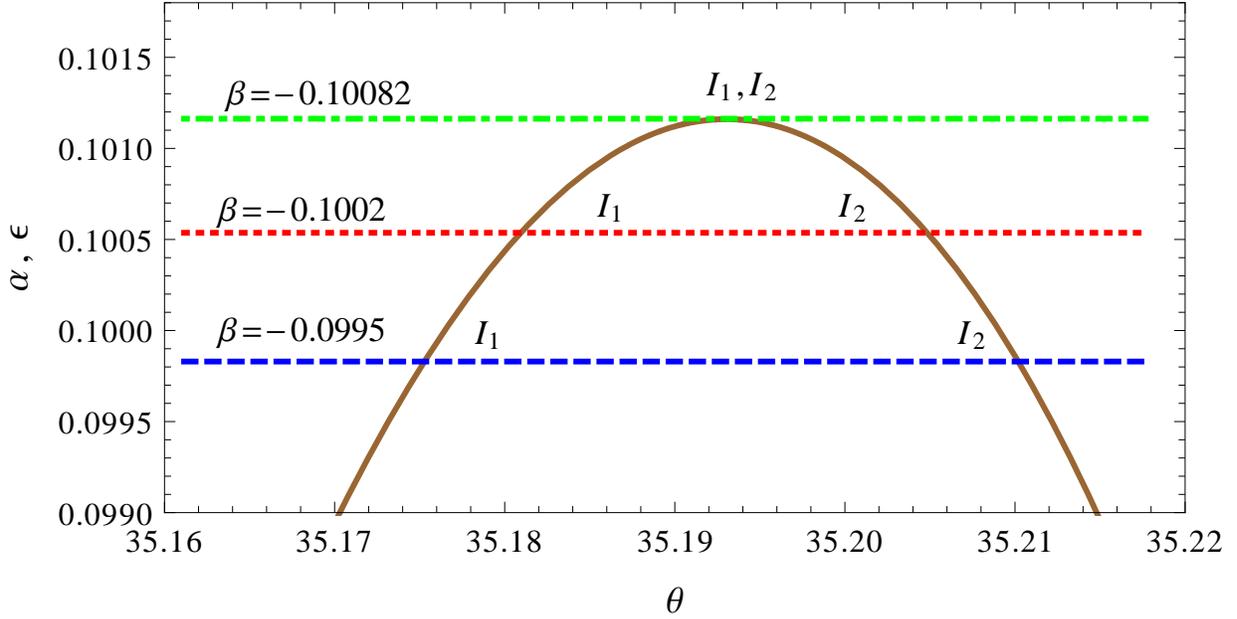}
\caption{\label{rc}
$\alpha=\frac{1}{2}\left(\tan\theta+\tan(\hat{\alpha}_0 -\theta)\right)$ (brown 
curve) and $\epsilon=\tan\theta-\tan\beta$ (green, red and blue dotted curves)
are plotted against $\theta$ expressed in micro-arcsecond for M=1.65. Lens 
equation is satisfied when  $\alpha$ and $\epsilon$ intersect and the image 
location $\theta$ can be read off from the coordinates of the intersection 
point. For the source location $\beta=-0.0995$ the two images denoted by $I_1$ 
and $I_2$ are located at $\theta_1=35.174$ and $\theta_2=35.21$. As we decrease 
$\beta$ images move closer and for $\beta=-0.1002$, images are located at 
$\theta_1=35.18$ and $\theta_2=35.204$. Images move towards each other further 
as we decrease $\beta$ and for $\beta=-0.10082$ both images coincide and a 
single degenerate images is located at $\theta=35.193$. This stands for the 
radial caustic.}
\end{center}
\end{figure}

\begin{figure}
\begin{center}
\includegraphics[width=1\textwidth]{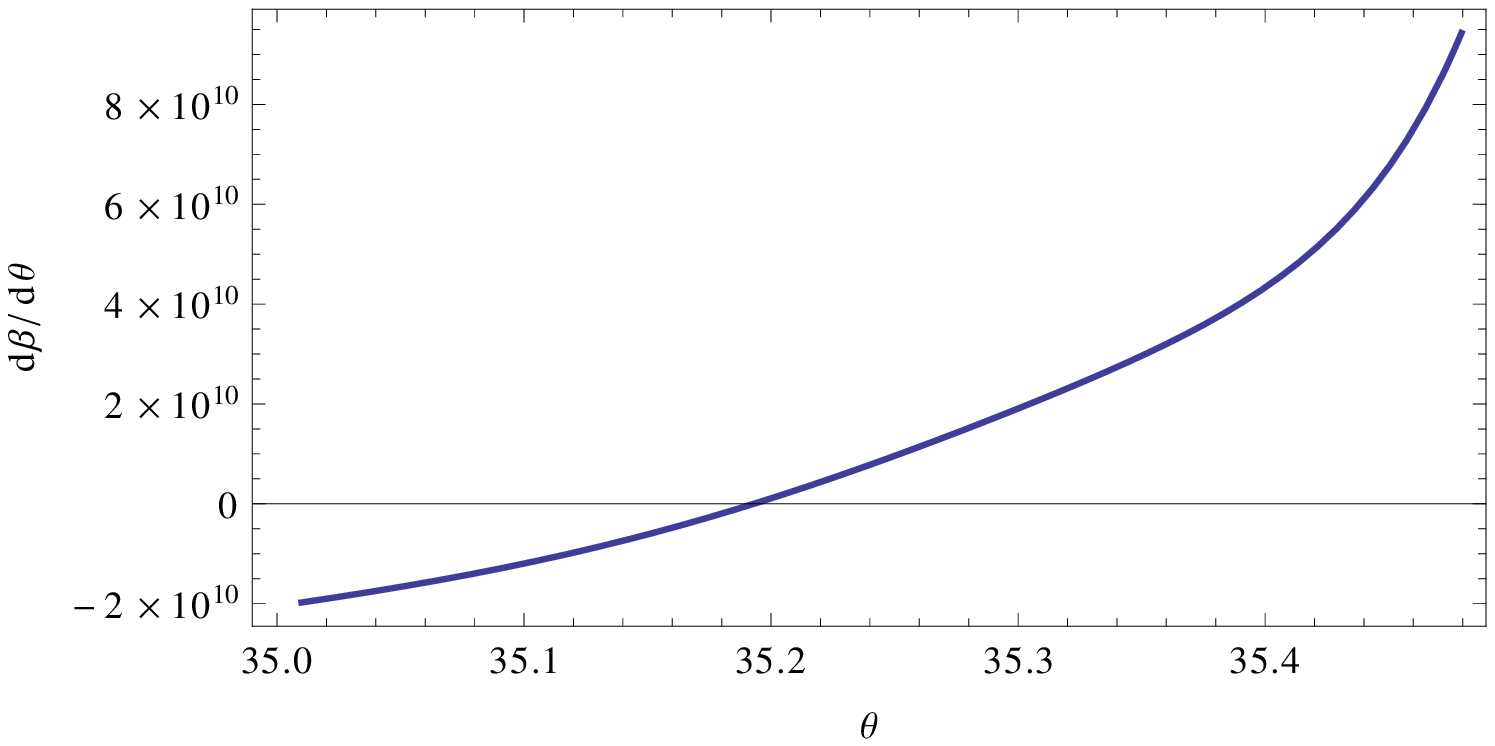}
\caption{\label{rc1}
$d\beta/d\theta$ is plotted against $\theta$ expressed in microarcsecond for the
parameter value $M=1.65$. The curve intersects x-axis at $\theta=35.193$ where 
the map from image plane to source plane is degenerate. This is the radial 
caustic.}
\end{center}
\end{figure}
 
We now deal with the final case where the parameter $M$ is less than the 
critical value $M_{*}=\sqrt{27/8}$. Here the effective potential increases 
monotonically as we decrease $\rho$ as shown in the Fig.(\ref{nph}). It does 
not admit extremum and thus there are no photon spheres on the equatorial plane. Ingoing light rays can 
turn back from every possible value of $\rho$ depending on the value of the 
impact parameter. The divergence of the deflection angle we encountered in the 
sections IV and V, was intimately related to the presence of the photon 
sphere. But now in this case in the absence of the photon sphere the deflection 
angle is finite for all values of $\rho_0$. It is zero for $\rho_0=0$ and 
$\rho_0 \rightarrow \infty$ and it admits maximum at some intermediate location 
of the turning point. 

We now deal with a specific value of the parameter $M=1.65$ and carry out a 
numerical computation. We numerically solve the lens equation and obtain the 
location of images in near-aligned configuration for source location 
$\beta=0.075$ radian. Rest of the parameters are chosen in a way explained in 
section III. Images that appear on the right side of the optic axis are located 
at $\theta=28.07$, $\theta=34.93$, $\theta=35.4$ and $\theta=36.58$, whereas 
the images that appear on the left side of the optic axis are located at 
$\theta=-28.78$, $\theta=-35.12$, $\theta=-35.28$ and $\theta=-36.42$. All image 
locations are expressed in microarcsecond. Thus we witness a formation of 
finitely many, discrete, well-separated images in the absence of the 
photon sphere. 

We now vary $\beta$ so as we intend to look for the radial caustic. 
In the Fig.\ref{rc} we have plotted $\alpha=\frac{1}{2}\left(\tan\theta+
\tan(\hat{\alpha}_0 -\theta)\right)$ and $\epsilon=\tan\theta-\tan\beta$ 
against $\theta$. The lens equation is satisfied at the intersection of two 
curves and the image location for a given source location can be read off from 
the coordinates of the intersection point. For a fixed value of $\beta$ there 
are two intersection points as it can be seen in the figure which corresponds 
to the two images. Images move closer as we decrease the value of $\beta$. 
Two images are located at $\theta_1=35.174$ and $\theta_2=35.21$ for 
$\beta=-0.0995$ and for smaller value of $\beta=-0.1002$ images are located 
closer to one another at $\theta_1=35.18$ and $\theta_2=35.204$. Eventually two 
images coincide at $\theta=35.193$ for $\beta=-0.10082$ as shown in the figure. 
The merger of two images separated from another in the radial direction with 
respect to the lens is referred to as radial caustic. Thus the radial caustic is 
admitted in the case where no photon sphere is present. As we can see from 
Fig.\ref{rc1}, the map from image plane to source plane is degenerate at the 
location of radial caustic as $d\beta/d\theta=0$. 

The radial caustic has never been observed before in any other investigation of 
relativistic gravitational lensing so far. Radial caustic is different from the tangential 
caustic which refers to the appearance of the Einstein ring or the merger of two images 
for a given source location that are separated from one another in tangential sense with respect 
to the lens.
Thus we discover a novel future in 
case of the binary black hole when separation between the black holes is 
moderately large. For a distance between black hole that is too large i.e. when 
$M<1.1$ the deflection angle is small compared to $2\pi$ and thus no 
relativistic images are formed.

\section{Conclusions and discussion}

In this paper we studied the gravitational lensing by binary black holes. In the
absence of any simple exact solution depicting the realistic binary black hole 
scenario, we resort to the use of Majumdar-Papapetrou metric which is the 
simplest multi-black hole solution is general relativity. We consider a 
scenario where the Majumdar-Papapetrou solution represents two equal mass 
maximally charged non-rotating black holes in equilibrium separated from one 
another by some distance. We focused on the case where the source of 
light and observers were located on the symmetry plane midway between the black 
holes and the light was allowed to move on the plane. The central point on the 
symmetry plane which is also the point midway between the black hole acts a 
gravitational lens. The source, observer and the lens are assumed to be in the 
near-aligned configuration. This allows us to employ the standard techniques
developed to study the gravitational lensing in the spherically symmetric 
spacetimes on the equatorial plane. 

The photon sphere which is the circular photon orbit plays a crucial role in the
relativistic gravitational lensing. The dihole spacetime can admit multiple 
photon spheres depending on the parameter $M$ in the Majumdar-Papapetrou di-hole 
metric which is the ratio of the mass and distance between the black holes. 
When $M$ is greater than the critical value given by $M_{*}=\sqrt{27/8}$, i.e. 
when for a fixed mass two black holes are located close enough, two photon 
spheres are present in the plane midway between the black holes. For the 
critical value $M=M_{*}=\sqrt{27/8}$, two photon spheres coincide and we have a 
single degenerate photon sphere located at $\rho=\sqrt{5}$. As we further 
increase the 
the parameter $M$ beyond the critical value, i.e. when the distance between the 
black holes for a given mass is large enough, no photon spheres are present on 
the symmetry plane. The gravitational lensing signature is qualitatively 
different depending on whether the photon spheres are present or not. 

Only one photon sphere is present in case of the single black hole. The light 
rays can have turning points outside the photon sphere, whereas the light rays 
that enter the photon sphere do not admit turning point and enter the black 
hole. Thus the images are formed only due to the light rays that turn back 
outside the photon sphere, all of which lie beyond certain critical angular 
radius with respect to the optic axis. There is a dark region 
below the critical radius. The situation is different in case of dihole due to 
the presence of the second photon sphere. The effective potential now 
turns upwards again below the second inner photon sphere and thus the 
light rays which enter the outer photon sphere can turn back again. A new set of 
infinitely many images are formed due to the light rays which enter photon 
sphere and turn back which are formed below the critical radius and fill up the 
dark void region which is present in case of the single black hole. Thus the 
pattern of images is drastically different in the presence of second inner 
photon sphere. The pattern of images in the presence of a single degenerate 
photon sphere is quite similar to the two photon sphere case, but the deflection 
angle shows power law divergence as opposed to logarithmic divergence in case of 
two photon spheres. 

In the absence of the photon sphere the deflection angle remains finite 
and we witness a formation of finitely many discrete well-separated image. 
Interestingly we also find the presence of radial caustic i.e. as we change the 
source location, the two images which are separated from one another in the 
radial direction with respect to the lens, approach one another and eventually 
merge together to form a single image. The map from image plane to source plane 
is degenerate at that instant. The radial caustic has never been reported 
before in any of the investigations so far related to the relativistic 
gravitational lensing. Therefore we discover a novel feature in the study of 
binary black holes when the distance between the two black holes is above a 
threshold value. 

The system consisting of two black holes nearby and a source of light in 
the vicinity could be found in the astrophysical context. During the binary black hole 
inspiral two black holes approach one another as the system loses energy due to the emission 
of the gravitational waves and the conditions conducive to the realization of radial caustic 
and the new relativistic images in the region between the two black holes can potentially be met.
The analysis carried out in this paper may possibly capture the some of the features related to the 
gravitational lensing signature of the realistic binary black hole system. 
New relativistic images and radial caustic can introduce additional features in the
microlensing light curve of binary black holes. As gravitational lensing signature could be
quite peculiar, it will allow us to identify the binary black hole system using electromagnetic 
observations. It would make it possible to infer the location of the binary black 
hole system on the sky and also possibly distance to the source. Thus it will 
reduce the number of parameters in the search for gravitational waves from 
binary black holes in the interferometric detectors using the technique of 
matched filtering. This suggests that our investigation may have 
implications for gravitational wave data analysis. We note that 
the static di-hole system that is analyzed in this work is a good
toy-model from which interesting physics can be studied as a proof of
concept and one would have to resort to a numerical relativity simulations of the 
binary black hole to come up with the concrete predictions \cite{Bohn}.

\section*{Acknowledgements}
M.P. would like to thank S. Sahu for discussions related to two photon spheres.

\end{document}